\documentclass[prd,superscriptaddress,showpacs,nofootinbib,amsmath,amssymb,aps,11pt]{revtex4}

\usepackage{bm}
\usepackage{amsfonts}
\usepackage{latexsym}
\usepackage[latin1]{inputenc}
\usepackage{graphicx}
\usepackage{amsmath}
\usepackage{palatino}
\usepackage{mathpazo}
\usepackage{booktabs}
\usepackage{dcolumn}

\usepackage{ulem}

\def\jnl@style{\it}
\def\aaref@jnl#1{{\jnl@style#1}}

\def\aaref@jnl#1{{\jnl@style#1}}

\def\aj{\aaref@jnl{AJ}}                   
\def\apj{\aaref@jnl{ApJ}}                 
\def\apjl{\aaref@jnl{ApJ}}                
\def\apjs{\aaref@jnl{ApJS}}               
\def\apss{\aaref@jnl{Ap\&SS}}             
\def\aap{\aaref@jnl{A\&A}}                
\def\aapr{\aaref@jnl{A\&A~Rev.}}          
\def\aaps{\aaref@jnl{A\&AS}}              
\def\mnras{\aaref@jnl{Mon.~Not.~Roy.~Astron.~Soc.}}             
\def\prd{\aaref@jnl{Phys.~Rev.~D}}        
\def\prc{\aaref@jnl{Phys.~Rev.~C}}  
\def\prl{\aaref@jnl{Phys.~Rev.~Lett.}}    
\def\qjras{\aaref@jnl{QJRAS}}             
\def\skytel{\aaref@jnl{S\&T}}             
\def\ssr{\aaref@jnl{Space~Sci.~Rev.}}     
\def\zap{\aaref@jnl{ZAp}}                 
\def\nat{\aaref@jnl{Nature}}              
\def\aplett{\aaref@jnl{Astrophys.~Lett.}} 
\def\apspr{\aaref@jnl{Astrophys.~Space~Phys.~Res.}} 
\def\physrep{\aaref@jnl{Phys.~Rep.}}      
\def\physscr{\aaref@jnl{Phys.~Scr}}       
\def\commat{\aaref@jnl{Comm.~Math.~Phys.}}              
\def\science{\aaref@jnl{Science}}               
\def\cqg{\aaref@jnl{Classical Quant.~Grav.}}            
\def\jpcs{\aaref@jnl{JPCS}}                                     
\def\ijmpd{\aaref@jnl{Int.~J.~Mod.~Phys.~D}}                    
\def\grg{\aaref@jnl{Gen.~Relat.~Gravit.}}               
\def\rpp{\aaref@jnl{Rep.~Prog.~Phys.}}          
\def\npa{\aaref@jnl{Nucl.~Phys.~A}}        
\def\lrr{\aaref@jnl{Living Rev.~Rel.}}                   
\def\jcap{\aaref@jnl{J.~Cosmology Astropart.~Phys.}}    
\def\rmp{\aaref@jnl{Rev.~Mod.~Phys.}}   

\allowdisplaybreaks[1]

\begin{document}

\title{Axial perturbations of hairy Gauss-Bonnet black holes with massive self-interacting scalar field }

\author{Kalin V. Staykov}
\email{kstaykov@phys.uni-sofia.bg}
\affiliation{Department of Theoretical Physics, Faculty of Physics, Sofia University, Sofia 1164, Bulgaria}

\author{Jose Luis Bl\'azquez-Salcedo}
\email{jlblaz01@ucm.es}
\affiliation{Departamento de F\'isica Te\'orica and IPARCOS, Universidad Complutense de Madrid, E-28040 Madrid, Spain}

\author{Daniela D. Doneva}
\email{daniela.doneva@uni-tuebingen.de}
\affiliation{Theoretical Astrophysics, Eberhard Karls University of T\"ubingen, T\"ubingen 72076, Germany}
\affiliation{INRNE - Bulgarian Academy of Sciences, 1784  Sofia, Bulgaria}

\author{Jutta Kunz}
\email{jutta.kunz@uni-oldenburg.de}
\affiliation{Institut f\"ur  Physik, Universit\"at Oldenburg, Postfach 2503, D-26111 Oldenburg, Germany}

\author{Petya Nedkova}
\email{pnedkova@phys.uni-sofia.bg}
\affiliation{Department of Theoretical Physics, Faculty of Physics, Sofia University, Sofia 1164, Bulgaria}

\author{Stoytcho S. Yazadjiev}
\email{yazad@phys.uni-sofia.bg}
\affiliation{Theoretical Astrophysics, Eberhard Karls University of T\"ubingen, T\"ubingen 72076, Germany}
\affiliation{Department of Theoretical Physics, Faculty of Physics, Sofia University, Sofia 1164, Bulgaria}
\affiliation{Institute of Mathematics and Informatics, 	Bulgarian Academy of Sciences, 	Acad. G. Bonchev St. 8, Sofia 1113, Bulgaria}


\begin{abstract}
	We study the axial quasinormal modes of hairy black holes in Gauss-Bonnet gravity with massive self-interacting scalar field. Two coupling functions of the scalar field to the Gauss-Bonnet invariant are adopted with one of them leading to black hole scalarization. The axial perturbations are studied via time evolution of the perturbation equation, and the effect of the scalar field mass and the self-interaction constant on the oscillation frequency and  damping time is examined. We study as well the effect of nonzero scalar field potential on the critical point at which the perturbation equation loses hyperbolicity in the case of black hole scalarization. The results show that the non-zero scalar field potential extends the range of parameters where such loss of hyperbolicity is observed thus shrinking the region of stable black hole existence. This will have an important effect on the nonlinear dynamical simulation studies in massive scalar Gauss-Bonnet gravity.
\end{abstract}


\maketitle

\section{Introduction}

Alternative theories of gravity which in the weak field regime pass all observational tests, but allow for large deviations in the strong field regime, are considered viable alternatives to general relativity (GR) and extensively studied for decades. The recent detections \cite{Abbott2016,Abbott2019,Abbott2019a,Barack2019,Abbott2020} of gravitational waves (GW) gave additional boost to the interest in such theories. The GW observations presented us with the prospect of testing strong  gravity via direct observations of the most compact objects in the Universe, black holes (BH) and neuron stars, in an entirely new observational channel, independent of the electromagnetic spectrum.

A specific class of viable and well motivated alternative theories of gravity  are the so-called extended scalar-tensor theories  \cite{Berti2015a}. One particular subclass that attracted considerable attention recently is the Gauss-Bonnet (GB) gravity where the scalar field is coupled to the GB invariant. The GB theory has the advantage that its field equations are of second order, like in GR, therefore there are no Ostrogradski instabilities or ghosts. In addition, black hole solutions with nontrivial scalar hair exist in GB gravity -- the presence of the Gauss-Bonnet invariant allows for the no-hair theorems to be circumvented (for a review see \cite{Berti2015}).

Hairy black holes in a particular shift-symmetric class of GB theory, the so-called Einstein-dilaton-Gauss-Bonnet (EdGB)  gravity where the scalar field is coupled to the GB invariant via a linear function of the scalar field or an exponent of a linear function, have been studied for many years \cite{Kanti1996,Torii1997,Guo2008,Pani2009,Pani2011d,Kleihaus2011,Ayzenberg2014,Ayzenberg2014a,Maselli2015a,Kleihaus2014a,Kleihaus2016,Cunha2017,Zhang2017}. Recently new scalarized black hole solutions were proven to exist for a different set of coupling functions being quadratic in the scalar field \cite{Doneva2018,Silva2018,Antoniou2018}. We will call this $Z_2$ symmetric class of theories scalar-Gauss-Bonnet (sGB) gravity. A key feature is that for a specific set of parameters, the Schwarzschild solution becomes unstable and new branches of scalarized black hole solutions emerge. The source of the scalar field is not some kind of a matter field but instead the curvature of the spacetime itself, thus we can name it curvature induced scalarization \cite{Doneva2018,Silva2018,Antoniou2018,Antoniou2018a,Doneva:2018rou,Minamitsuji2019,Silva2019,Brihaye2019,Myung2019,Hod2019,Cunha:2019dwb,Collodel:2019kkx}. Multiple such branches of scalarized solutions exist and they can be labeled by the number of nodes that the scalar field has. It turned out, though, that only the branch having no nodes can be stable \cite{BlazquezSalcedo2018}. Later, black hole solutions with curvature induced scalarization were constructed for sGB theories with massive and self-interacting massive scalar field \cite{Macedo2019,Doneva2019,Bakopoulos2020} as well as scalarized black hole with curvature induced scalarization \cite{Dima:2020yac,Hod:2020jjy,Doneva:2020nbb,Herdeiro:2020wei,Berti:2020kgk,Doneva:2020kfv}.

The quasi-normal modes (QNMs) of EdGB black holes are considered in multiple papers \cite{BlazquezSalcedo2016,BlazquezSalcedo2017,Konoplya2019,Zinhailo2019} (see \cite{Blazquez-Salcedo:2018pxo} for a review). The QNMs of the sGB scalarized black holes were quite thoroughly examined in the last  couple of years as well. In \cite{BlazquezSalcedo2018} the authors study the radial stability of the  sGB solutions obtained in \cite{Doneva2018}. They show that only black holes from the fundamental branch (the one with nodeless scalar field) are stable againts radial perturbations. The perturbation equation loses hyperbolicity for small masses at some critical point but all solutions between this point and the bifurcation point are radially stable. On the other hand, all Schwarzschild black hole solutions in sGB gravity with $r_H$ smaller than the bifurcation point of the fundamental branch are unstable. Therefore, there is some value of $r_H$ under which no stable black hole solutions exist, neither scalarized nor unscalarized. The stability analysis of those black hole solutions was later extended to the case of axial perturbations \cite{BlazquezSalcedo2020}. The authors of  \cite{BlazquezSalcedo2020} report that the axial perturbation equation loses hyperbolicity at a critical point with slightly higher $r_H$ compared to the radial case.  The black hole solutions between this critical radius and the bifurcation point, though, are shown to be stable under axial perturbations. Therefore, the branch with axially stable solutions is slightly shorter compared to the radially stable one. The authors of \cite{BlazquezSalcedo2020} observe small deviations in the oscillation frequencies and damping times compared to the Schwarzschild case. The linear stability analysis of the sGB black hole solutions  \cite{Doneva2018} was completed in \cite{BlazquezSalcedo2020a} where the authors study the polar perturbations of the stable models from \cite{BlazquezSalcedo2020}. It is shown that all studied models are stable under polar perturbations. In addition, models with various multipole numbers ($l = 0, 1, 2$) were studied. In all cases there are significant deviations in the oscillation frequencies and damping times. Breaking of the isospectrality between the axial and the polar modes was reported as well. 

In the present paper we study the axial perturbations of black holes in EdGB gravity with massive self-interacting scalar field, and we concentrate on two coupling functions: linear coupling (EdGB black holes) and coupling function which allows for scalarized black hole solutions with curvature induced scalarization (sGB black holes). The axial perturbations we study by performing time evolution of the time-dependent perturbation equation. In the axial case, the scalar field perturbations are zero, and the scalar field potential does not present itself explicitly in the perturbation equation, hence the master perturbation equation will be the same as the one reported in \cite{BlazquezSalcedo2020}.

The paper is structured as follows: In Section II we present the main features of GB theory and the master equation for the axial perturbations. Section III we devoted to the numerical results for both coupling functions. In Section IV  special attention is paid to loss of hyperbolicity for the sGB theory and the effect of nonzero scalar field potential on it. The paper ends with Conclusions.

\section{Mathematical background}

\subsection{Gauss-Bonnet gravity}

In the general case, Gauss-Bonnet theories are described by the action

\begin{eqnarray}\label{GBA}
S=&&\frac{1}{16\pi}\int d^4x \sqrt{-g} 
\Big[R - 2\nabla_\mu \varphi \nabla^\mu \varphi - V(\varphi) 
+ \lambda^2 f(\varphi){\cal R}^2_{GB} \Big] ,\label{eq:quadratic}
\end{eqnarray}
where $R$ is the Ricci scalar with respect to a spacetime metric $g_{\mu\nu}$, and the Gauss-Bonnet invariant is defined by ${\cal R}^2_{GB}=R^2 - 4 R_{\mu\nu} R^{\mu\nu} + R_{\mu\nu\alpha\beta}R^{\mu\nu\alpha\beta}$, where $R_{\mu\nu}$ is the Ricci tensor and $R_{\mu\nu\alpha\beta}$ is the Riemann tensor. The scalar field  potential  $V(\varphi)$ and coupling function $f(\varphi)$ depend only on the scalar field $\varphi$. The Gauss-Bonnet coupling constant $\lambda$ has dimension of $length$. The field equations derived by the above action are the following
\begin{eqnarray}
&&R_{\mu\nu}- \frac{1}{2}R g_{\mu\nu} + \Gamma_{\mu\nu}= 2\nabla_\mu\varphi\nabla_\nu\varphi -  g_{\mu\nu} \nabla_\alpha\varphi \nabla^\alpha\varphi - \frac{1}{2} g_{\mu\nu}V(\varphi), \label{FE1}\\
&&\nabla_\alpha\nabla^\alpha\varphi= \frac{1}{4} \frac{dV(\varphi)}{d\varphi} -  \frac{\lambda^2}{4} \frac{df(\varphi)}{d\varphi} {\cal R}^2_{GB}, \label{FE2}
\end{eqnarray}
where  $\nabla_{\mu}$ is the covariant derivative with respect to the spacetime metric $g_{\mu\nu}$ and  $\Gamma_{\mu\nu}$ is defined by 
\begin{eqnarray}
&&\Gamma_{\mu\nu}= - R(\nabla_\mu\Psi_{\nu} + \nabla_\nu\Psi_{\mu} ) - 4\nabla^\alpha\Psi_{\alpha}\left(R_{\mu\nu} - \frac{1}{2}R g_{\mu\nu}\right) + 
4R_{\mu\alpha}\nabla^\alpha\Psi_{\nu} + 4R_{\nu\alpha}\nabla^\alpha\Psi_{\mu} \nonumber \\ 
&& - 4 g_{\mu\nu} R^{\alpha\beta}\nabla_\alpha\Psi_{\beta} 
+ \,  4 R^{\beta}_{\;\mu\alpha\nu}\nabla^\alpha\Psi_{\beta} 
\end{eqnarray}  
with 
\begin{eqnarray}
\Psi_{\mu}= \lambda^2 \frac{df(\varphi)}{d\varphi}\nabla_\mu\varphi .
\end{eqnarray}
In the present study we shall focus on a massive potential with fourth order self-interacting term, namely   
\begin{equation}\label{eq:Potential}
V(\varphi)=2 m_\varphi^2\varphi^2 + \Lambda \varphi^4,
\end{equation}
where $m_\varphi$ is the mass of the scalar field, and $\Lambda$ is the self-interaction constant.

We will be interested in static and spherically symmetric metric and scalar field configurations, therefore we will use the following general ansatz for the background metric
\begin{eqnarray}\label{eq:metric_BG}
ds^2= - e^{2\mu(r)}dt^2 + e^{2\nu(r)} dr^2
+ r^2 (d\theta^2 + \sin^2\theta d\phi^2 ) \ .
\end{eqnarray}
For more details on the GB theory as well as the explicit form of the reduced field equations for the above metric, and the appropriate boundary conditions, we refer the reader to \cite{Doneva2019}.

In the present study we will present results for the two most prominent classes of Gauss-Bonnet gravity, namely the shift symmetric EdGB theory and the $Z_2$ symmetric sGB theories in correlation with \cite{Doneva2019}. In the former case the coupling function is linear with respect to the scalar field   
\begin{equation} \label{eq:coupling_function_lin}
f(\varphi) = \varphi,
\end{equation} 
and for the latter theory allowing for spontaneous scalarization we can choose
\begin{equation} \label{eq:coupling_function_scal}
f(\varphi)=  \frac{1}{2 \beta} \left(1- e^{-\beta \varphi^2}\right). \ 
\end{equation}
Even though this coupling function is not the simplest one that can lead to scalarization it is perhaps among the best choices for numerical simulations and studying astrophysical implications because it leads to stable well behaved branches of black hole solutions \cite{BlazquezSalcedo2018}. We will fix the value of $\beta = 12$  in accordance with \cite{Doneva2019}.

The system of reduced field equations has to be solved simultaneously with the appropriate boundary conditions that in our case come from the requirement for regularity at the black hole horizon and asymptotic flatness at infinity. Details can be found for example in \cite{Doneva2018} and here we will mention only one prominent feature. Something characteristic for Gauss-Bonnet theories is that the regularity condition at the horizon (or at the stellar center in case of neutron stars for example) can be violated leading to termination of the hairy black hole branches. The exact form of the condition has a complicated form and for pure GB gravity with a nonzero scalar field potential it can be found in \cite{Macedo2019,Doneva2019}. The above choice of coupling function leading to scalarization \eqref{eq:coupling_function_scal} with $\beta=12$ is actually adjusted in such a way that scalarized black hole solutions exist all the way from the bifurcation point to zero black hole mass. Such adjustment is not possible, though, for sGB gravity with coupling \eqref{eq:coupling_function_lin} and thus the branches are terminated at some finite black hole mass as we will see below.

\subsection{Axial Perturbations}

The axial perturbations of the scalar field are zero by definition, thus the nonvanishing scalar field potential will not appear in the perturbation equation, which will be the same as the one derived in \cite{BlazquezSalcedo2020}. Instead the influence of the QNM spectrum will be indirect through the altered BH solutions in GB gravity with nonzero scalar field potential. In this section we will skip the derivation and we will present only the final result. 

The wave-type master equation for the axial perturbations has the form  

\begin{eqnarray}\label{master_eq}
\frac{\partial^2\Psi}{\partial \tilde{r}_{*}^2}&+&\left[\frac{1}{2\mathcal{S}_0}\frac{\partial^2\mathcal{S}_0}{\partial \tilde{r}_{*}^2}-\frac{3}{4\mathcal{S}_0^2}\left(\frac{\partial\mathcal{S}_0}{\partial \tilde{r}_{*}}\right)^2-\frac{1}{r\mathcal{S}_0}\frac{\partial\mathcal{S}_0}{\partial \tilde{r}^{*}}e^{\mu_0-\nu_0}-\frac{e^{2\mu_0-2\nu_0}}{r^2}(2-r(\mu_0'-\nu_0')) \right. \nonumber \\[3mm]
&&\left. -(l-1)(l+2)\,\frac{e^{2\mu_0}\mathcal{S}_0}{r^2\mathcal{W}_0}\right]\Psi =\frac{\mathcal{S}_0}{\mathcal{P}_0}\partial^2_{t}\Psi \ ,
\end{eqnarray} 
where $\Psi$ is a proper combination of the metric perturbations, $l$ is a positive integer, and $\tilde{r}_{*}$ is the Regge-Wheeler (tortoise) coordinate,
defined by $\frac{\partial}{\partial r}
=e^{\nu_0-\mu_0}\frac{\partial}{\partial \tilde{r}_{*}}$. $\mathcal{P}_0(r)$, $\mathcal{W}_0(r)$, and $\mathcal{S}_0(r)$ are auxiliary functions of the background quantities defined as follows:
\begin{eqnarray}
&&\mathcal{P}_0(r)=1-4\lambda^2\,\frac{df(\varphi_0)}{d\varphi_0}\,\mu_0'\,\varphi_0'\, e^{-2\nu_0} \ ,  \\[3mm]
&&\mathcal{W}_0(r)=1-4\lambda^2\frac{df(\varphi_0)}{d\varphi_0}\,\frac{\varphi_0'}{r}\, e^{-2\nu_0} \ ,  \\[3mm]
&&\mathcal{S}_0(r)=1-4\lambda^2\frac{d^2f(\varphi_0)}{d\varphi_0^2}(\varphi_0')^2\,e^{-2\nu_0} - 4\lambda^2\frac{df(\varphi_0)}{d\varphi_0}\,\varphi_0''\,e^{-2\nu_0} +  4\lambda^2\frac{df(\varphi_0)}{d\varphi_0}\,\nu_0'\,\varphi_0'\,e^{-2\nu_0}.
\end{eqnarray}
In the above equations the derivative with respect to the radial coordinate is denoted by $()'$, and the background scalar field and metric functions are denoted with subscript zero.

For the purpose of our study we will keep the perturbation equation in its present time-dependent form. If the QNMs are to be studied as an eigenvalue problem, Eq.~($\ref{master_eq}$) should be transformed in time independent Sch\"odinger-like form. This is not in the scope of the present paper, however, we refer the interested reader to \cite{BlazquezSalcedo2020}.

For Eq.~($\ref{master_eq}$) to describe QNMs, appropriate boundary conditions should be imposed, namely we should have a purely outgoing wave at infinity, and purely ingoing wave at the horizon:
\begin{eqnarray}
\Psi \xrightarrow[r \to \infty]{} e^{-i\omega(t-\tilde{r}_{*})} \ , \nonumber \\
\Psi \xrightarrow[r \to r_{\rm H}]{} e^{-i\omega(t+\tilde{r}_{*})} \ . \label{eq:BC_in_outgoing}
\end{eqnarray}
Those conditions can be imposed in more convenient differential form when solving the time-dependent problem: 
\begin{eqnarray}
&&{\rm horizon:}\;\;\; \partial_t \Psi - \partial_{\tilde{r}_{*}} \Psi =0 \ , \nonumber \\
&&{\rm infinity \; :}\;\;\;   \partial_t \Psi + \partial_{\tilde{r}_{*}} \Psi =0 \ .
\end{eqnarray}

\section{Background numerical solutions}
In the present paper we would like to study the effect of a scalar field potential, having the form of Eq. \eqref{eq:Potential} with both a massive and a self-interacting term, on the axial QNM frequencies of black holes in GB gravity. The background black hole solutions for $m_\varphi>0$ and $\Lambda=0$ were already obtained in \cite{Doneva2019}. Here we will review briefly these results and extend them to the case of nonzero~$\Lambda$. The values of the scalar field mass and the self-interaction constant are chosen in order to span a range of cases. All quantities and parameters are rescaled with the Gauss-Bonnet coupling constant $\lambda$ in the proper way.

\subsection{Shift symmetric EdGB black holes}

\begin{figure}[]
	\centering
	\includegraphics[width=0.45\textwidth]{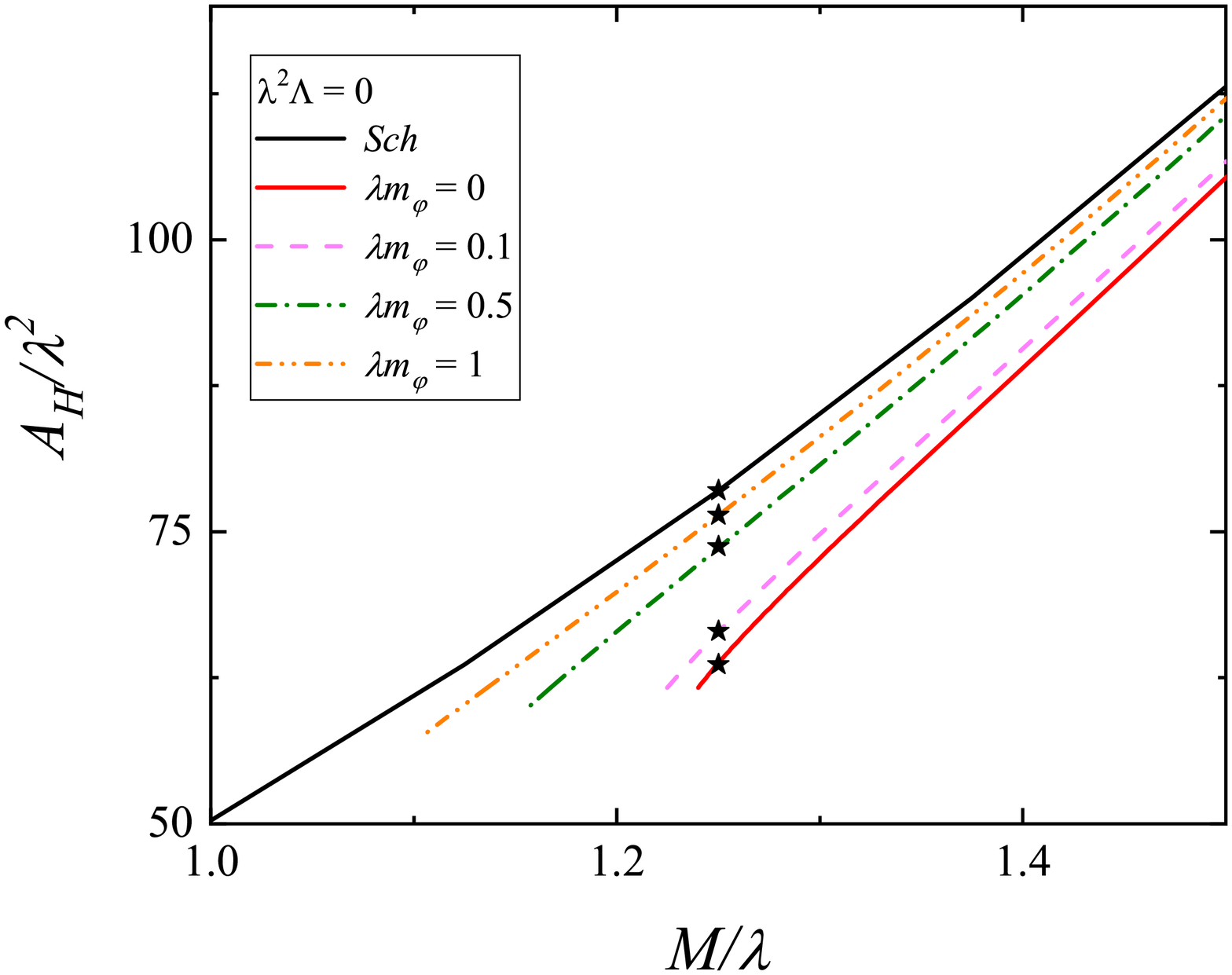}
	\includegraphics[width=0.45\textwidth]{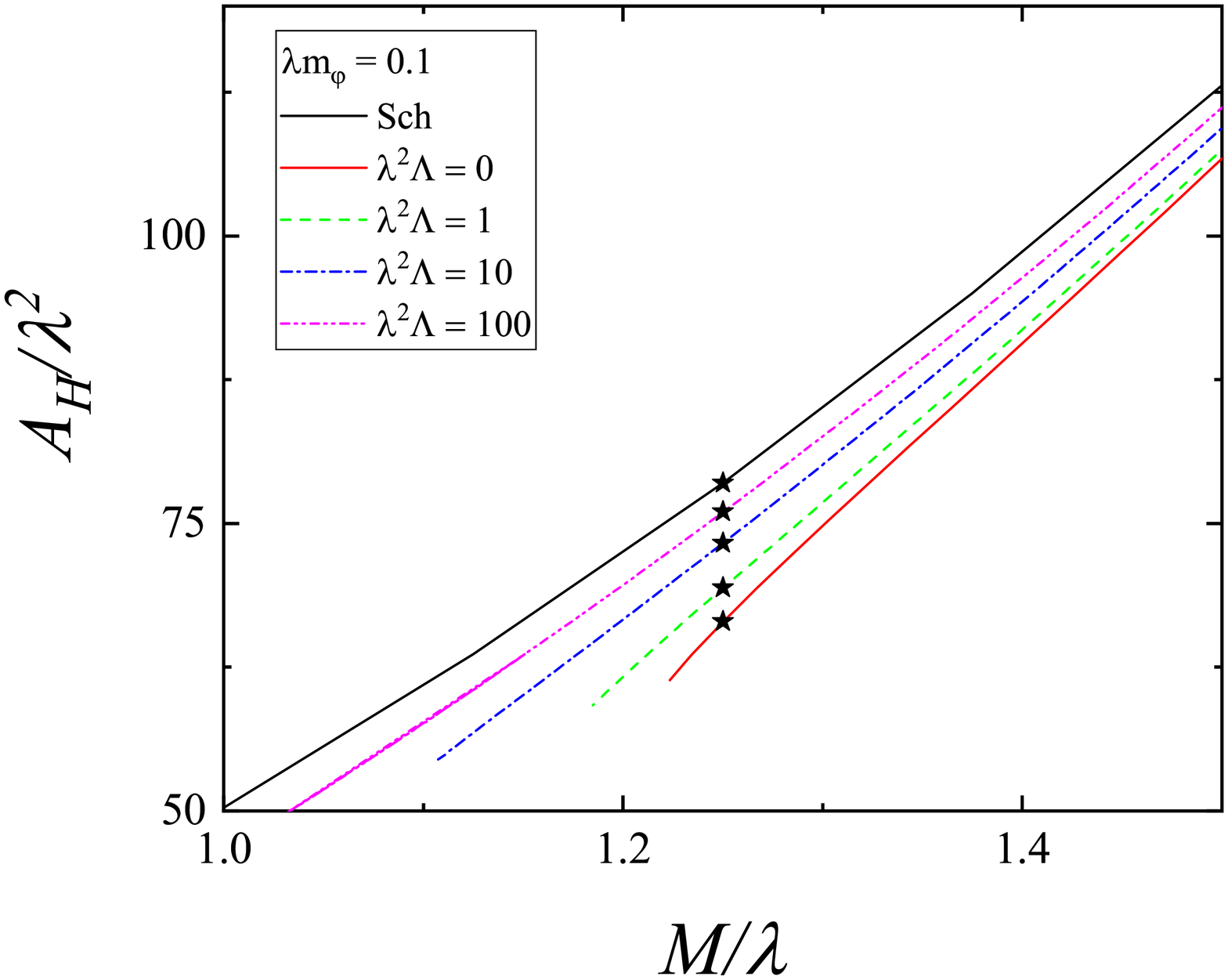}
	\caption{The area of the horizon as a function of the black hole mass for EdGB theory with linear coupling \eqref{eq:coupling_function_lin}. \textit{Left} Pure massive theory with different values of the scalar field mass. \textit{Right} Massive theory with self-interaction -- fixed scalar field mass and different values of the self-interaction constant. The models for which we calculate the QNMs in the following section are marked with black asterisk}
	\label{Fig:bg_lin}
\end{figure}

Firstly, we study shift-symmetric EdGB black holes with linear coupling function given by eq. \eqref{eq:coupling_function_lin}. As a first step, we extend the solutions presented in \cite{Doneva2019} by adding a self-interacting term in the potential and obtain black hole solutions for a wide range of combinations between the dimensionless mass of the scalar field $\lambda m_{\varphi}$ and the dimensionless self-interaction constant $\lambda^2 \Lambda$. In Fig. \ref{Fig:bg_lin} we present the area of the black hole horizon as a function of the black hole mass. As one can see from the figures, for this coupling function there is a cutoff mass (or radius), below which no black hole solutions exist in the EdGB theory that is due to violation of the regularity condition at the black hole horizon  (see e.g. \cite{ Doneva2019}). In the \textit{left} panel the pure massive case without self-interaction is presented for different values of the scalar field mass. The maximal deviation from GR is for the pure GB case and it decreases in the massive case with the increase of the scalar field mass. In the \textit{right} panel we present the results for massive theory with self-interaction for fixed value of the scalar field mass and different values of the self-interaction constant $\lambda^2 \Lambda$. In this case the maximal deviation is for the pure massive theory and it decreases in the self-interaction case with the increase of $\lambda^2 \Lambda$. The black hole models for which we calculate the QNMS later in this work are marked with black asterisks.

\subsection{$Z_2$ symmetric scalarized sGB black holes}

\begin{figure}[]
	\centering
	\includegraphics[width=0.45\textwidth]{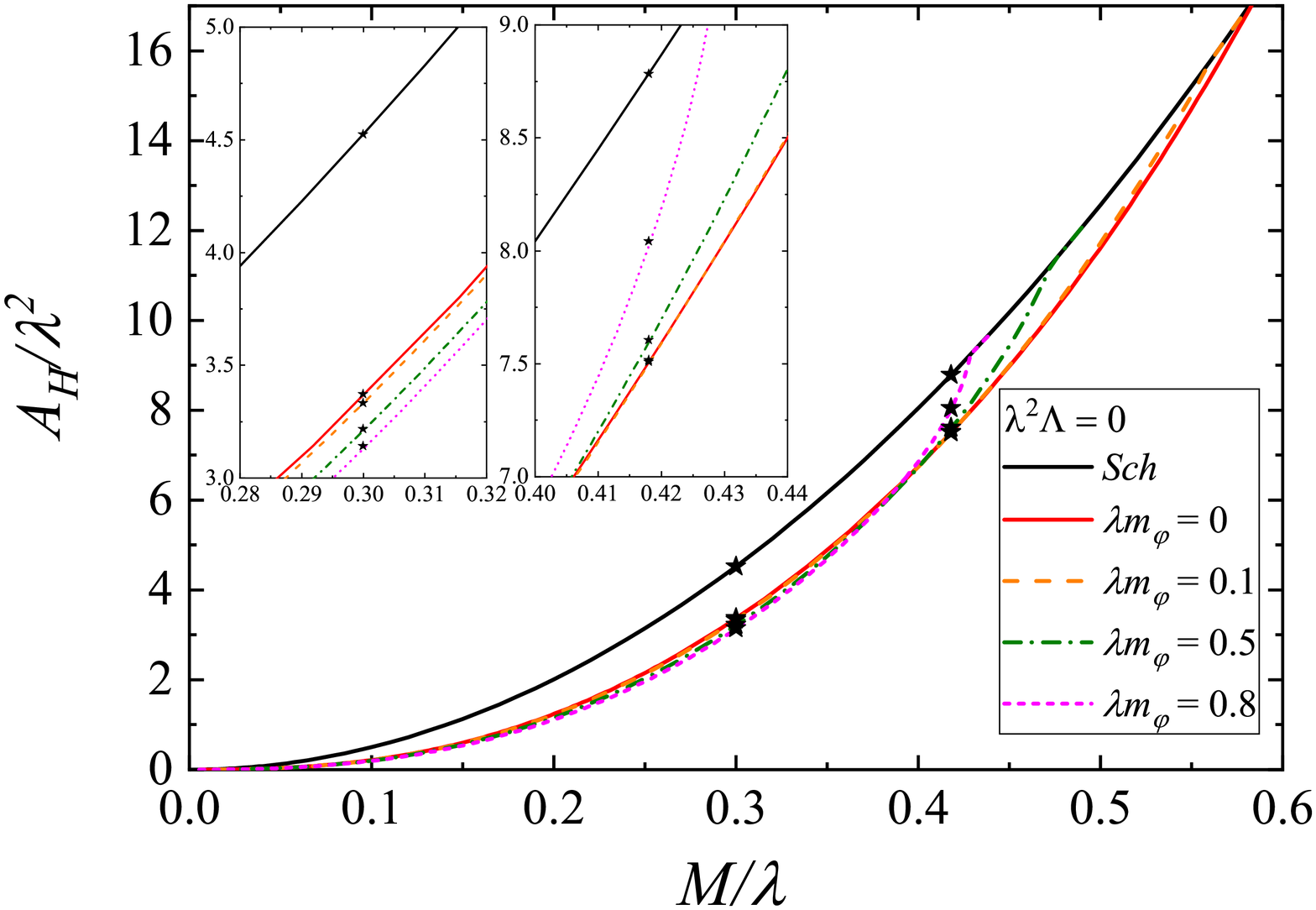}
	\includegraphics[width=0.45\textwidth]{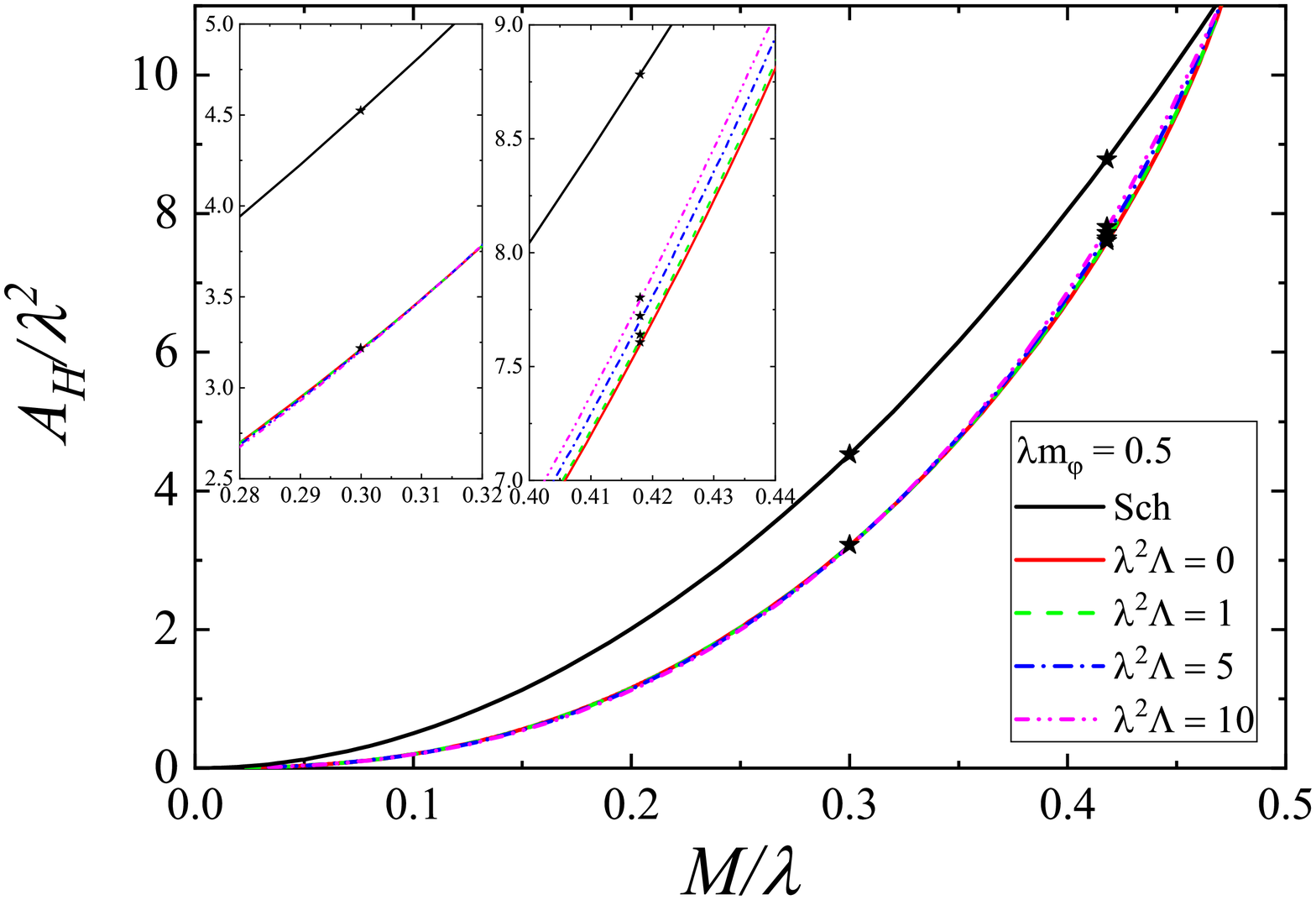}
	\caption{ The area of the horizon as a function of the black hole mass for sGB. \textit{Left} Pure massive theory with different values of the scalar field mass. \textit{Right} Massive theory with self-interaction -- fixed scalar field mass and different values of the self-interaction constant. The models for which we calculate the QNMs in the following section are marked with black asterisk}
	\label{Fig:bg_exp_Ah}
\end{figure}

\begin{figure}[]
	\centering
	\includegraphics[width=0.45\textwidth]{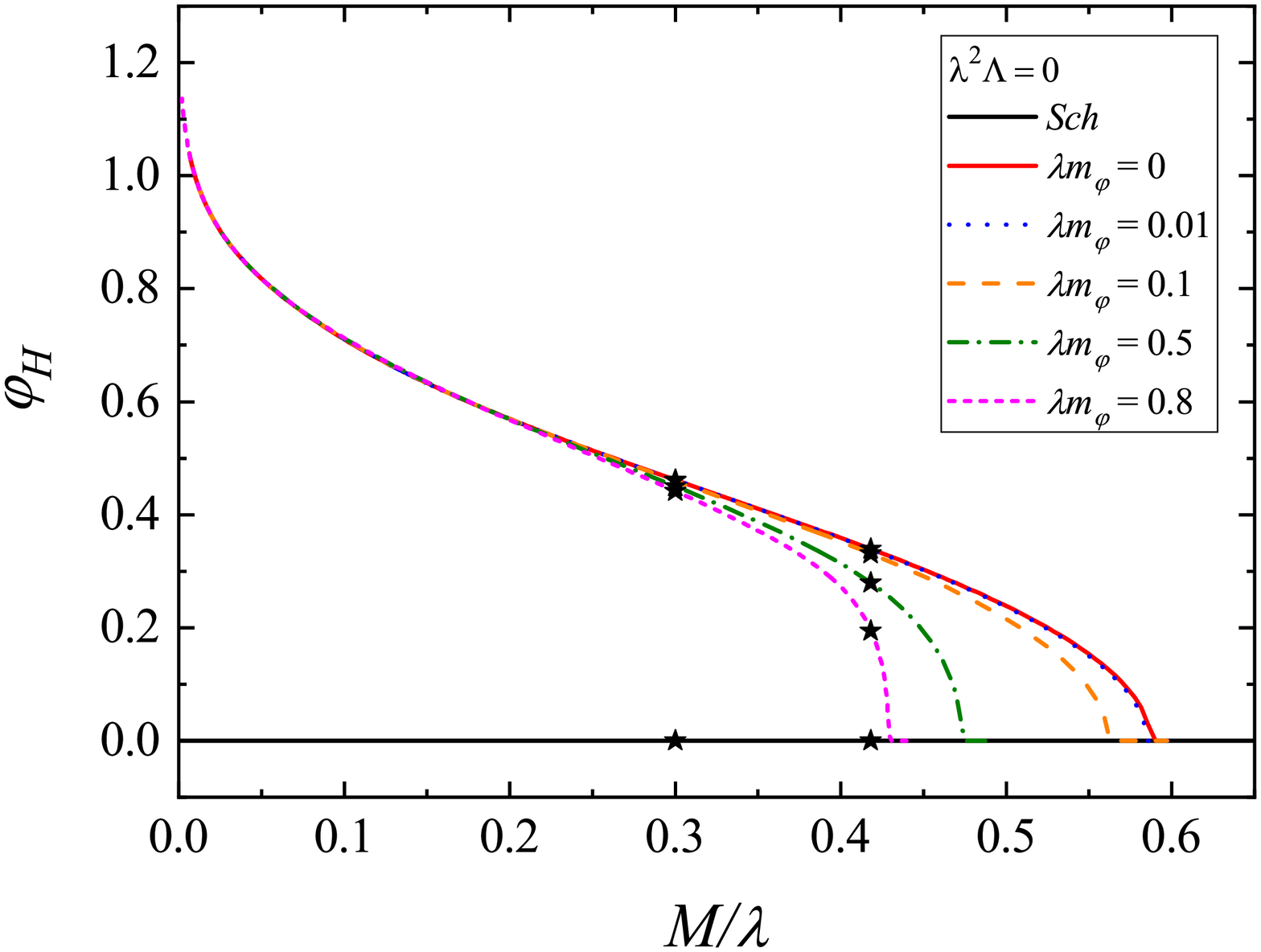}
	\includegraphics[width=0.45\textwidth]{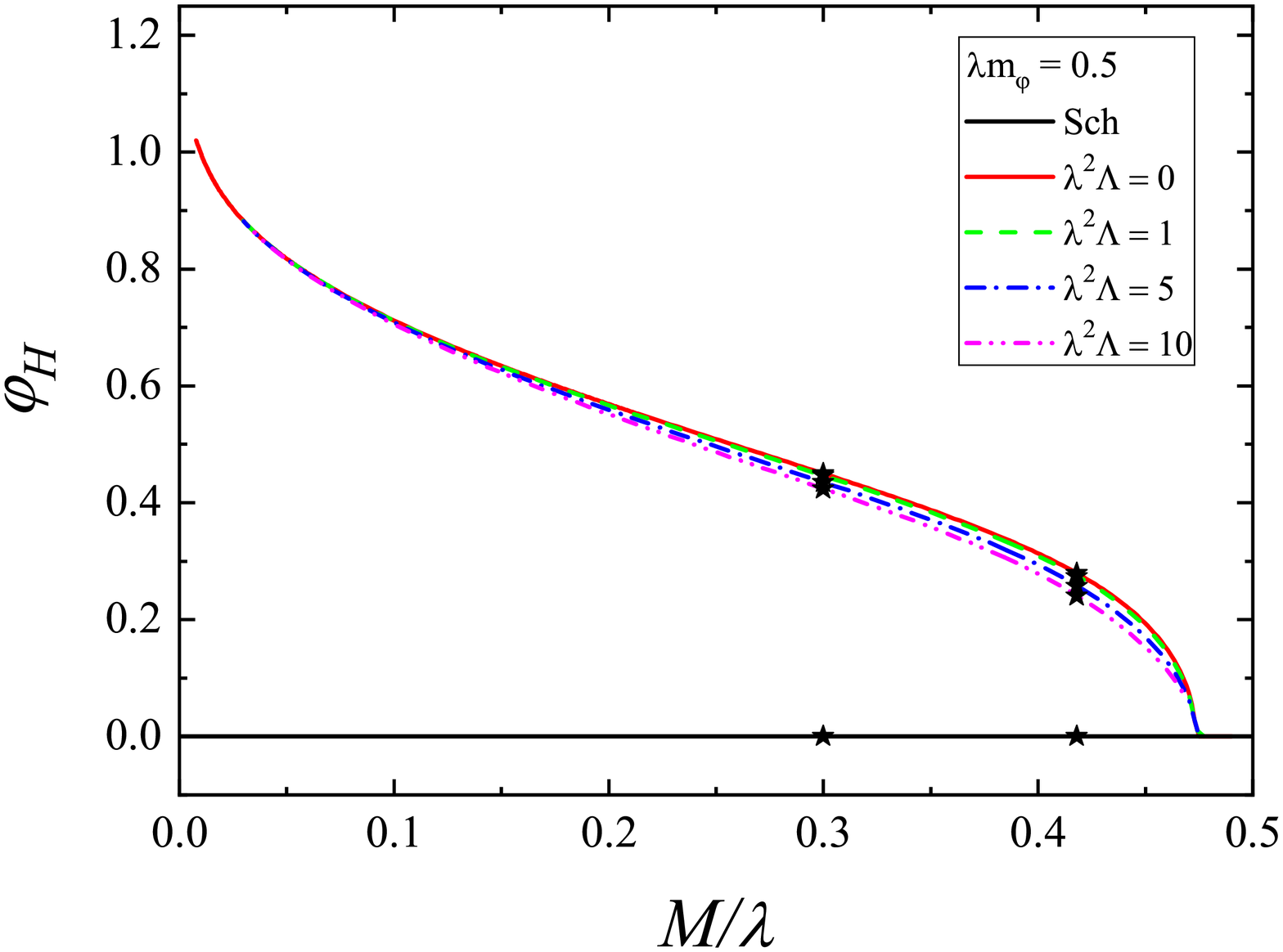}
	\caption{ The scalar field on the horizon as a function of the black hole mass for sGB with coupling function \eqref{eq:coupling_function_scal}. \textit{Left} Pure massive theory with different values of the scalar field mass. \textit{Right} Massive theory with self-interaction -- fixed scalar field mass and different values of the self-interaction constant. The models for which we calculate the QNMs in the following section are marked with black asterisk }
	\label{Fig:bg_exp_phi}
\end{figure}

We continue our study with sGB black holes with coupling function given by eq. \eqref{eq:coupling_function_scal} where we fix $\beta = 12$. The motivation behind this choice of $\beta$ is the following. As observed in \cite{Doneva:2018rou} the increase of $\beta$ leads on one hand to decrease of the deviations from GR but on the other, higher $\beta$ branches are behaving better from a numerical point of view and also in terms of existence of solutions. For example, there is a minimum $\beta$ below which no stable scalarized solutions can exist and for intermediate $\beta$ the branch can be terminated at some finite mass due to violation of the regularity condition at the horizon, similar to the EdGB case considered above. In that sense $\beta=12$ is a safe choice where such problems would not appear \cite{Doneva:2018rou}. Moreover, as the experience in \cite{Doneva2019}  shows, it is numerically easier to calculate scalarized black holes with massive scalar field for larger $\beta$ that is in general a challenge due to stiffness introduced with increasing $m_\varphi$ and  $\Lambda$. Since our goal is to explore a wide range of scalar field potential parameters we have decided to stay in the better behaved  $\beta$ range.

The results from \cite{Doneva2019} were extended for a potential with self-interacting term of the form (\ref{eq:Potential}) and in Fig. \ref{Fig:bg_exp_Ah} we present the area of the horizon as a function of the black hole mass for sGB black holes with curvature induced scalarization. In the \textit{left} panel the pure massive case is presented for different values of the scalar field mass. The major effect of the scalar field mass in this case is on the bifurcation point -- the increase of $\lambda m_{\varphi}$ moves the bifurcation point to lower black hole masses. For better visualization and distinction between the branches, in the \textit{left} panel of Fig. \ref{Fig:bg_exp_phi} we plot the scalar field on the horizon as a function of the black hole mass for the same black hole models. The effect of $\lambda m_{\varphi}$ is clear in this plot.

In the \textit{right} panel of Fig. \ref{Fig:bg_exp_Ah} we plot the self-interaction case for fixed value of the scalar field mass and different values of the self-interaction constant. The corresponding scalar field on the horizon is plotted in the \textit{right} panel of Fig. (\ref{Fig:bg_exp_phi}). As expected, the presence of the self-interaction term in the potential \eqref{eq:Potential} does not move the bifurcation point. However, the increase of $\lambda^2\Lambda$ suppresses the scalar field. The threshold value of $\lambda^2\Lambda$ that is allowed from the regularity condition at the horizon is dependent on $\lambda m_{\varphi}$ and in the figure we have chosen $\lambda^2\Lambda$ to be close to the maximum one for the highest scalar field mass we study. As it can be seen in the figure this allows for quite small deviation from the zero scalar field potential case.

\section{Black hole quasi-normal modes}
The quasi-normal modes are damped oscillations with complex frequency. The real part of that frequency is the proper oscillation frequency, and the imaginary part is equal to the inverse damping time. If those frequencies are required with high accuracy, solving the eigenvalue problem is the way to go. The time-evolution, though, has its benefits, such as its simplicity and the fact that we can study the actual response of a black hole to different types of perturbations. These results are good enough, however, for quantitative estimations of the effect of the scalar field potential which suits the purpose of the current study. 

As an initial data we used a Gaussian pulse and the resulting signal is extracted at some large distance away from the black hole horizon. The real and the imaginary parts of the oscillation frequency are not straightforward to be extracted from the oscillation profile, and the accuracy of the results will depend on the observed numbers of cycles before the appearance of the asymptotic tail. In the present paper the real part of the frequency is extracted with Fourier transformation of the signal, and the imaginary part is obtained by determining the slope of the oscillation profile in logarithmic scale.    

In this section we will proceed to calculating the QNMs frequencies and damping times for the models presented above. More specifically, we will focus on the models marked with asterisk in Figs. \ref{Fig:bg_lin} and \ref{Fig:bg_exp_Ah} separately for the two subclasses of GB gravity we consider. We will focus on the axial perturbations with $l=2$.

\subsection{EdGB black holes perturbations}

We will focus first on the EdGB black holes with linear coupling \eqref{eq:coupling_function_lin}. We should note that even though QNMs of EdGB black holes were already considered in the literature in the massless scalar field case \cite{BlazquezSalcedo2016,BlazquezSalcedo2017,Konoplya2019}, a different exponential form of the coupling was adopted in these studies. Thus QNMs with the coupling \eqref{eq:coupling_function_lin} have not been considered until now even in the case of zero scalar field potential. It is expected, though, that the results should be qualitatively similar to \cite{BlazquezSalcedo2016,BlazquezSalcedo2017,Konoplya2019}.

\begin{figure}[]
	\centering
	\includegraphics[width=0.45\textwidth]{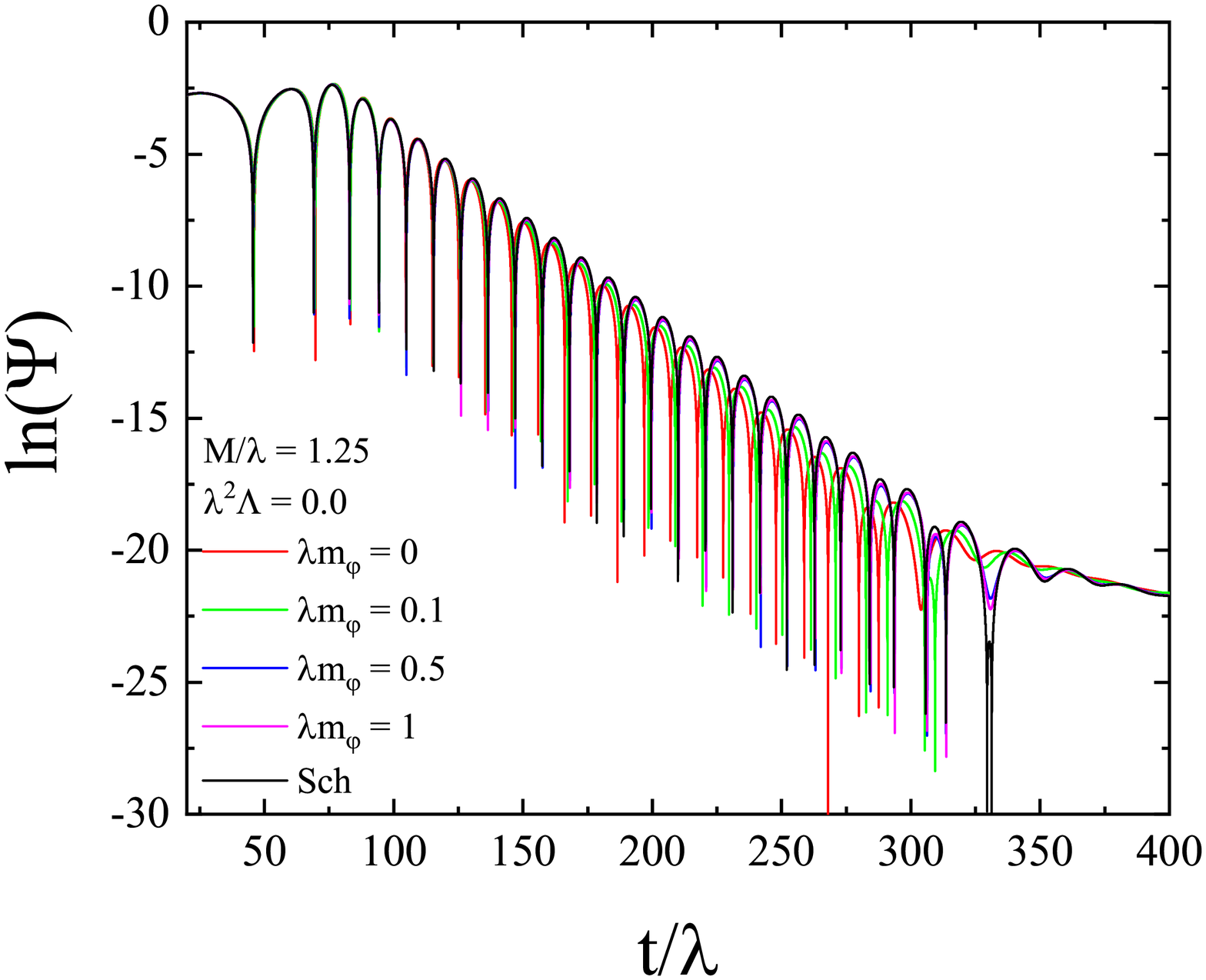}
	\includegraphics[width=0.45\textwidth]{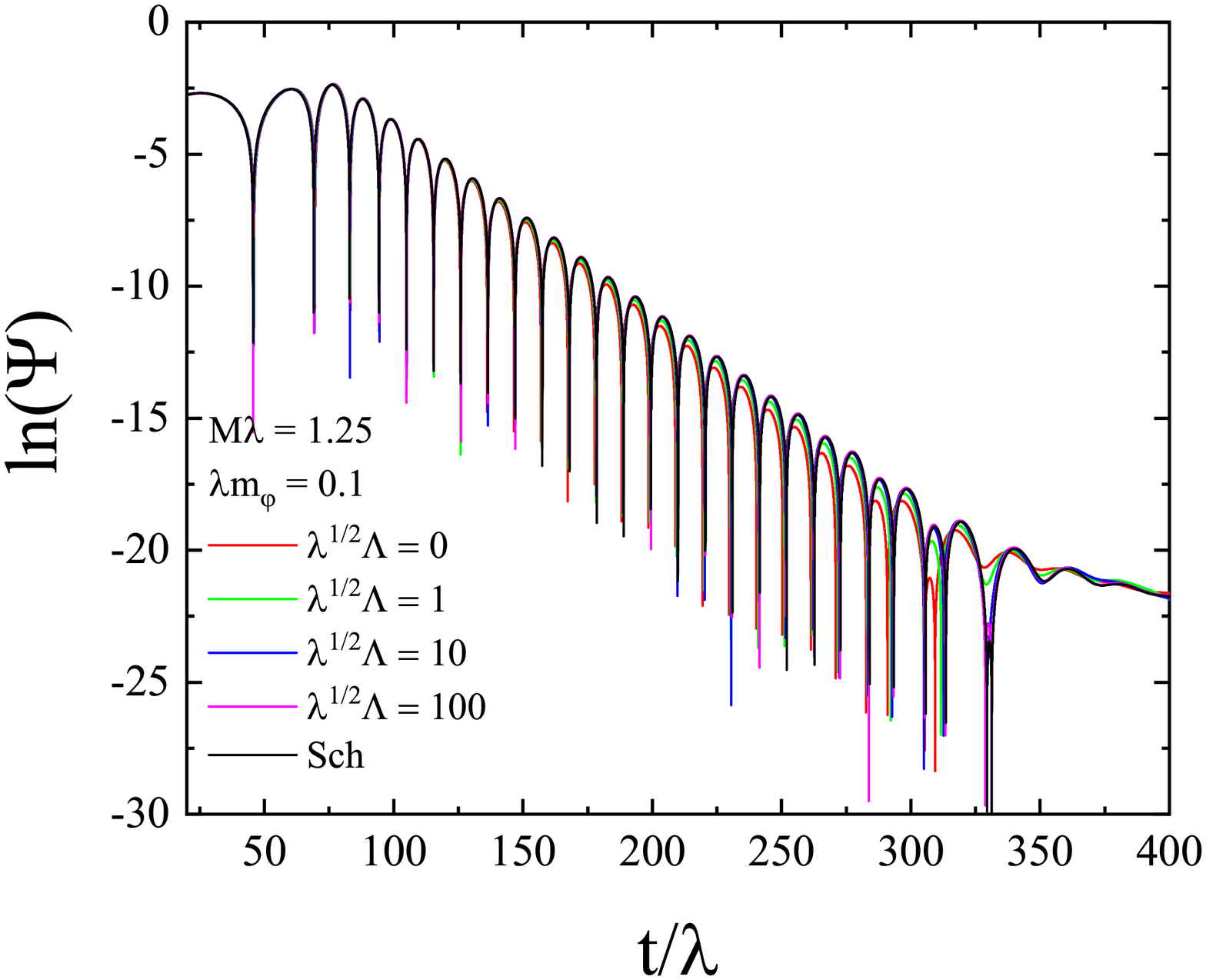}
	\caption{ Time evolution of the perturbation $\Psi$ in logarithmic scale for linear coupling \eqref{eq:coupling_function_lin} and black hole models with $ M/\lambda = 1.25$ and $l=2$. The signal is extracted at coordinate distance $r/\lambda = 50$. \textit{Left} Different values of the mass of the scalar field and no self-interaction. \textit{Right} Fixed mass of the scalar field, and different value of the self-interaction constant. On both figures the Schwarzschild solution is in black line. }
	\label{Fig:ln_H_rh2.25}
\end{figure}

The results presented below are for models with equal black hole mass $M/\lambda = 1.25$ (those marked with black asterisk in Fig. \ref{Fig:bg_lin}). They are chosen in such away, so that their mass is in the vicinity of the smallest allowed mass for all studied combinations of parameters. This secures deviations from GR close to the maximum allowed. In Fig. \ref{Fig:ln_H_rh2.25} we present, in logarithmic scale, the perturbation function $\Psi$ for these models. The massless EdGB case and the Schwarzschild solutions are presented for comparison. The results for the pure massive case (with no self-interaction) with different values for the mass of the scalar field are presented in the left panel.  From the slope of the graphs it is clear there is a small difference in the imaginary parts of the frequencies -- the pure EdGB case has the highest imaginary part and for the rest of the models it is slightly smaller, with the Schwarzschild one being the smallest. In terms of damping times, the Schwarzschild one is the longest, and the pure EdGB one -- the shortest. The real part of the frequency presents similar behaviour with the parameters.

In the right panel, the results for models with fixed mass of the scalar field and different values for the self-interaction constant are presented. A smaller value for the mass of the scalar field has been chosen in order for the deviations from GR not to be  suppressed predominantly by the massive term. In this case as well, the damping times and the oscillation frequencies are quite similar for all combinations of parameters. Similar to the case when  $\lambda m_{\varphi}$ is varied, here also both $\omega_R$ and $\omega_I$ decrease and approach the Schwarzschild values when $\lambda^2 \Lambda$ increases.

The numerical values for the real and the imaginary part of the frequency (in dimensionless unit) for all models plotted in Fig. \ref{Fig:ln_H_rh2.25} are given in Table \ref{Tbl:rh2.25}. As one can see, even though a clear difference between the background models in Fig. \ref{Fig:bg_lin} can be observed, this is not well evident in the QNM spectrum -- the oscillation frequency can differ only by roughly 2\% compared to Schwarzschild while the damping time of the modes changes a bit more of the order of 10\%. This suggests that it might be very difficult to distinguish between EdGB black holes and Schwarzschild in a real astrophysical scenario. One should keep in mind, though, that the main purpose of this paper is to examine the effect of nonzero scalar field potential. Thus we have taken one of the most popular EdGB couplings. It is well possible to be able to magnify this difference with a better choice of the coupling function.  

\begin{table}
	\caption{ The real and the imaginary part, in dimensionless units, of the frequency for the models presented in Fig. \ref{Fig:ln_H_rh2.25}.}
	\label{Tbl:rh2.25}
	\begin{center}
		\begin{tabular}{ | l | l | l | l |}
			\hline
			$\lambda m_{\varphi}$ & $\lambda^2\Lambda$ & $\lambda\omega_R$ & $\lambda\omega_I$  \\ \hline
			Sch & Sch & 0.0476 & $-0.0711 $ \\ \hline
			0 & 0 & 0.0490 & $-0.0767 $   \\ \hline
			0.1 & 0 & 0.0480 & $-0.0750 $   \\ \hline
			0.5 & 0 & 0.0475 & $-0.0719$   \\ \hline
			1.0 & 0 & 0.0475 & $-0.0717$  \\ \hline
			0.1 & 1 & 0.0478 & $-0.0727$   \\ \hline
			0.1 & 10 & 0.0478 & $-0.0713$   \\ \hline
			0.1 & 100 & 0.0477 & $-0.0710$   \\ \hline		
		\end{tabular}
	\end{center}
\end{table}

\subsection{sGB black hole perturbations}

Now we concentrate on the sGB black holes with curvature induced scalarization.
The presented results are for two black hole masses (the models marked with a black asterisk in Fig. \ref{Fig:bg_exp_Ah}), namely $M/\lambda = 0.3$ and $M/\lambda = 0.418$, and in both cases the signal is extracted at coordinate distance $r/\lambda = 13$. The larger mass is chosen to be in the vicinity of the bifurcation point for the branch with the highest mass of the scalar field (the leftmost  bifurcation point in Fig. \ref{Fig:bg_exp_Ah}). 

\begin{figure}[]
	\centering
	\includegraphics[width=0.45\textwidth]{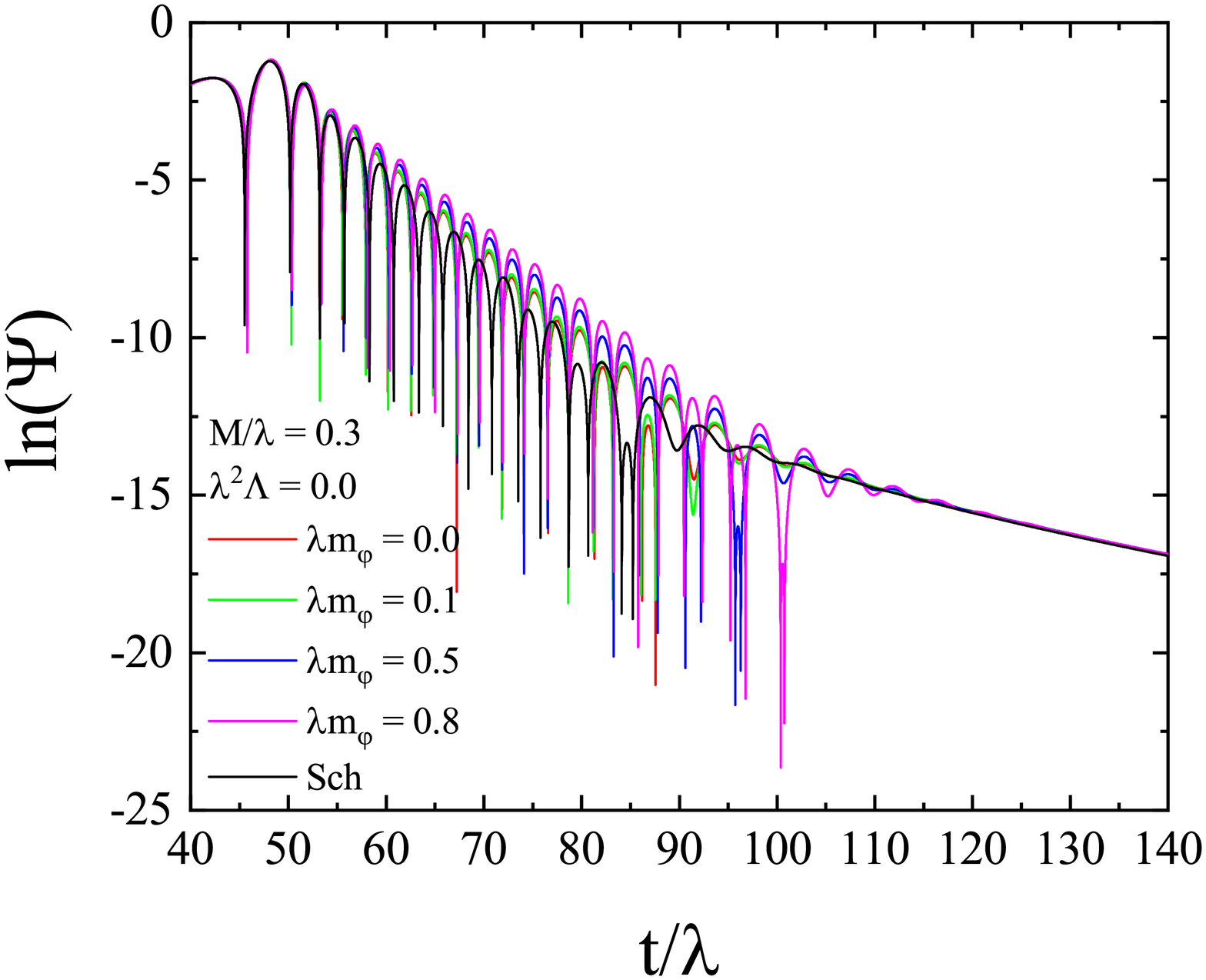}
	\includegraphics[width=0.45\textwidth]{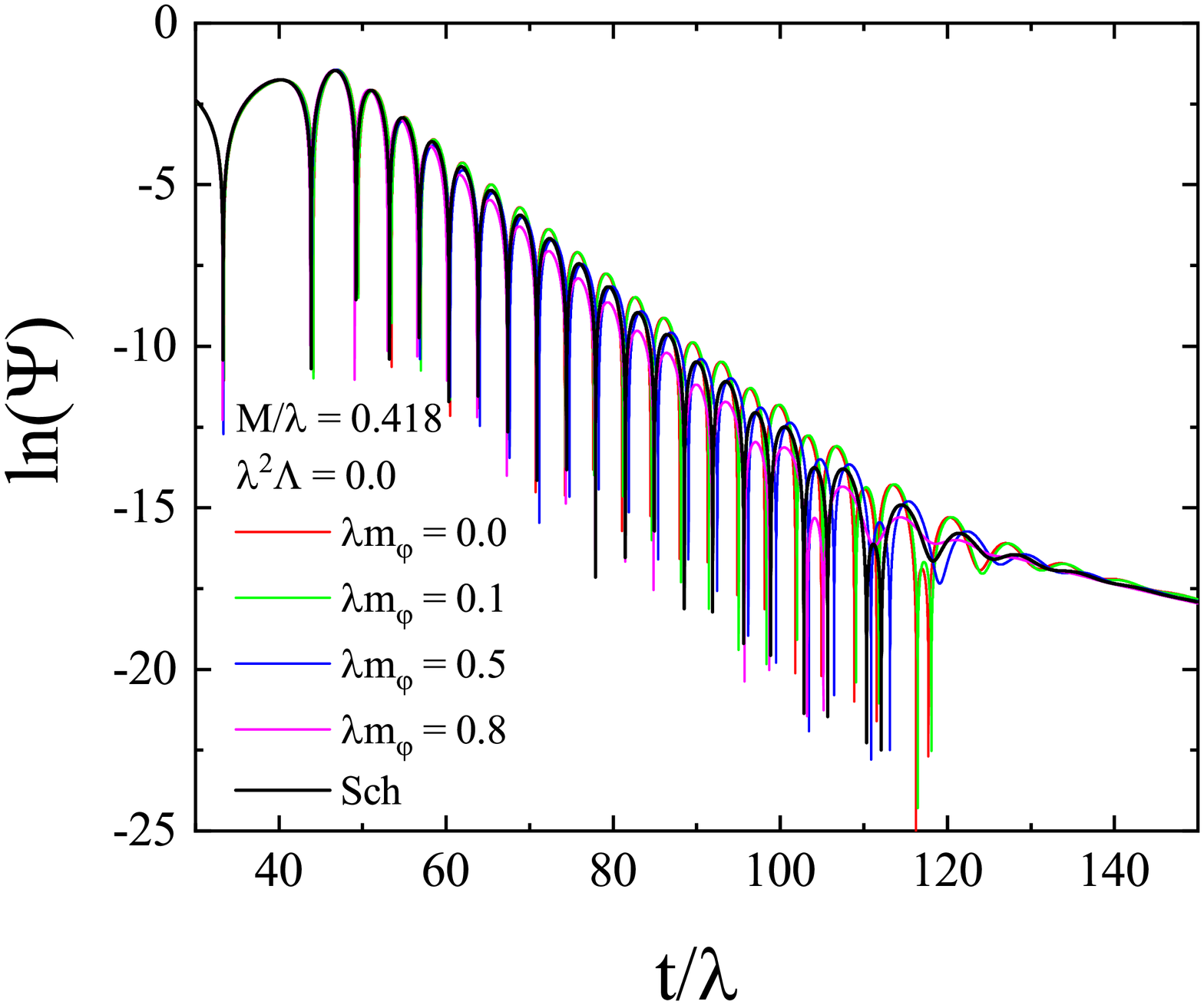}
	\caption{ Time evolution of the perturbation $\Psi$ in logarithmic scale for scalarized black holes (coupling function given by \eqref{eq:coupling_function_lin}) with different values of the mass of the scalar field and no self-interaction having $l=2$. The signal is extracted at coordinate distance $r/\lambda = 13$.  \textit{Left} Models with $M/\lambda = 0.3$ and \textit{Right} models with $M/\lambda = 0.418$. On both figures the Schwarzschild solution is in black line. }
	\label{Fig:ln_m_l00}
\end{figure}

\begin{figure}[]
	\centering
	\includegraphics[width=0.45\textwidth]{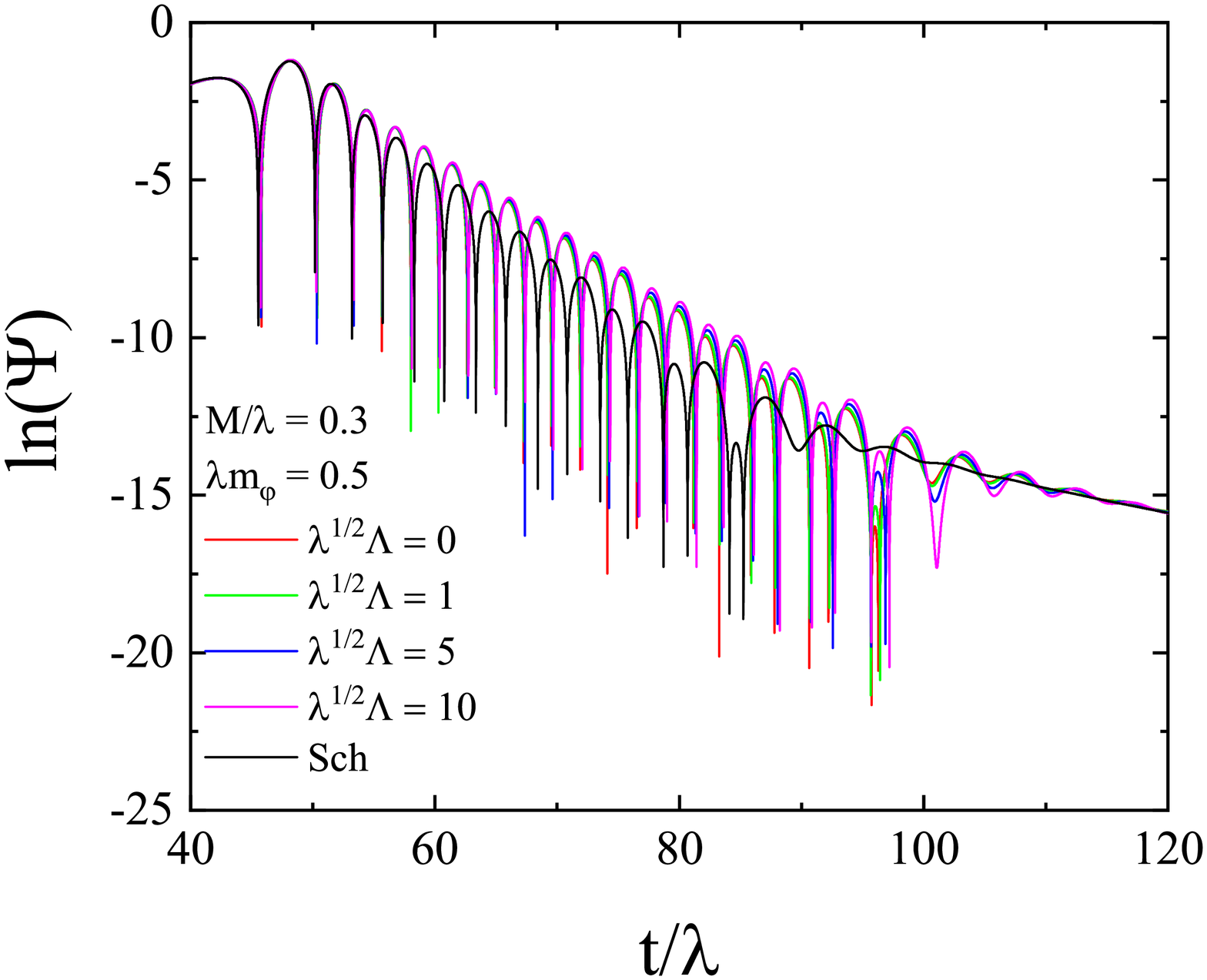}
	\includegraphics[width=0.45\textwidth]{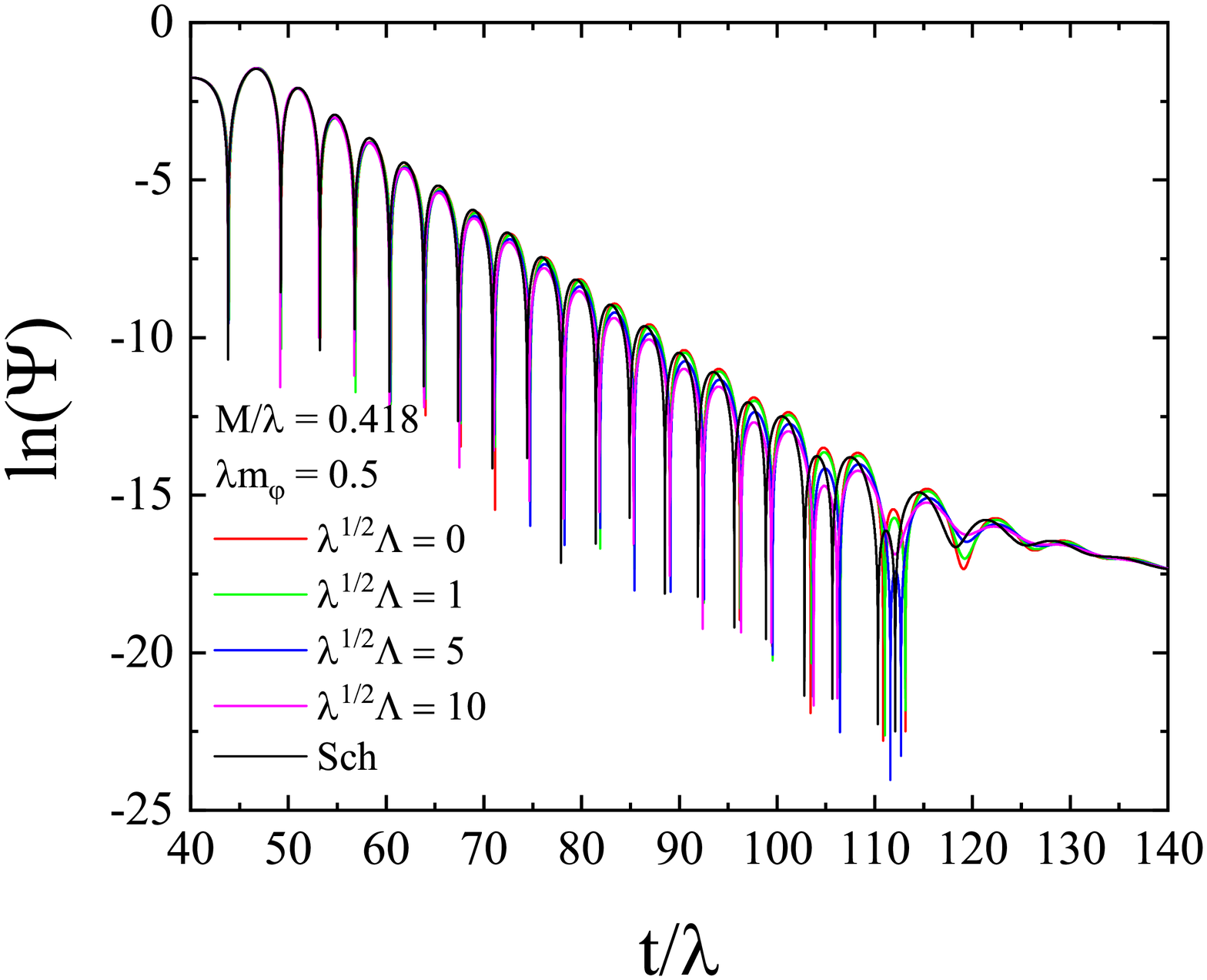}
	\caption{ Time evolution of the perturbation $\Psi$ in logarithmic scale for scalarized black holes (coupling function given by \eqref{eq:coupling_function_lin}) with fixed mass of the scalar field and different values of the self-interaction constant $\lambda^2\Lambda$. The signal is extracted at coordinate distance $r/\lambda = 13$.  \textit{Left} Models with $M/\lambda = 0.3$ and \textit{Right} models with $M/\lambda = 0.418$. On both figures the Schwarzschild solution is in black line and marked as Sch. }
	\label{Fig:ln_m05_l}
\end{figure}

In Fig. \ref{Fig:ln_m_l00} in logarithmic scale is presented the perturbation function $\Psi$ for the Schwarzschild solution and scalarized black hole solutions with different values of the scalar field mass and no self-interaction. Models with black hole mass  $M/\lambda = 0.3$ are presented in the left panel, and models with $M/\lambda = 0.418$ -- in the right one. For the $M/\lambda = 0.3$ case the highest imaginary part is for the Schwarzschild case, followed by the pure sGB case, and it continues to decrease with the increase of the mass of the field. Therefore, the Schwarzschild solution has the shortest damping time, and the solution with the maximal studied mass of the scalar field -- the longest. 
For the case with $M/\lambda = 0.418$ the model with the highest mass of the scalar field has imaginary part (damping time) quite similar to the Schwarzschild one, or even slightly higher (lower damping time), and the rest of the models have smaller but similar to one another imaginary parts. For all studied cases the oscillation frequency is lowest for the Schwarzschild case, but the deviations are as expected very small since the difference in the background black hole solutions are also small. The relevant numerical values are presented in Table \ref{Tbl:rh05} and Table \ref{Tbl:rh08}.  For models with fixed mass of the scalar field and different values of the self-interaction constant, similar behavior of the QNM frequencies is observed as evident from Fig. \ref{Fig:ln_m05_l}, however, the effect of the self-interaction constant is smaller.

\begin{table}
	\caption{ The real and the imaginary part, in dimensionless units, of the frequency for scalarized black holes (coupling function given by \eqref{eq:coupling_function_lin}) with mass $M/\lambda = 0.3$.}
	\label{Tbl:rh05}\textbf{}
	\begin{center}
		\begin{tabular}{ | l | l | l | l |}
			\hline
			$\lambda m_{\varphi}$ & $\lambda^2\Lambda$ & $\lambda\omega_R$ & $\lambda\omega_I$  \\ \hline
			Sch & Sch & 0.199 & $-0.302$  \\ \hline
			0 & 0 & 0.215 & $-0.283$  \\ \hline
			0.1 & 0 & 0.216 & $-0.280$ \\ \hline
			0.5 & 0 & 0.217 & $-0.258$  \\ \hline
			0.8 & 0 & 0.218 & $-0.242$  \\ \hline
			0.5 & 1 & 0.217 & $-0.255$  \\ \hline
			0.5 & 5 & 0.215 & $-0.250$  \\ \hline
			0.5 & 10 & 0.215 & $-0.243$  \\ \hline		
		\end{tabular}
	\end{center}
\end{table}

\begin{table}
	\caption{ The real and the imaginary part, in dimensionless units, of the frequency for scalarized black holes with mass $M/\lambda = 0.418$}
	\label{Tbl:rh08}\textbf{}
	\begin{center}
		\begin{tabular}{ | l | l | l | l |}
			\hline
			$\lambda m_{\varphi}$ & $\lambda^2\Lambda$ & $\lambda\omega_R$ & $\lambda\omega_I$  \\ \hline
			Sch & Sch & 0.142& $-0.213$  \\ \hline
			0 & 0 & 0.146 & $-0.202$  \\ \hline
			0.1 & 0 & 0.145 & $-0.201$  \\ \hline
			0.5 & 0 & 0.141 & $-0.206$ \\ \hline
			0.8 & 0 & 0.142 & $-0.227$ \\ \hline
			0.5 & 1 & 0.140 & $-0.208$  \\ \hline
			0.5 & 5 & 0.140 & $-0.214$ \\ \hline
			0.5 & 10 & 0.140 & $-0.222$  \\ \hline		
		\end{tabular}
	\end{center}
\end{table}

\subsection{Loss of hyperbolicity}

In this section we present  arguably the most important part of our study -- the effect of $\lambda m_{\varphi}$ and $\lambda^2 \Lambda$ on the critical point at which the  perturbation equation (\ref{master_eq}) loses hyperbolicity. As it turns out, even though the QNM frequencies discussed in the previous section are not strongly affected by the presence of a scalar field potential, the point where the hyperbolic character of the perturbation equation \eqref{master_eq} changes differs a lot as $m_\varphi$ and $\Lambda$ are varied.

The critical point at which the perturbation equation loses hyperbolicity can be determined by analysing the coefficient $S_0$. Loss of hyperbolicity occurs if $S_0$ as a function of the radial coordinate changes its sign. Hence the critical point is determined as the first black hole model for which such change in the sign is observed. This is in agreement with the results from time evolution within the numerical error.

We have thoroughly studied this problem for all combinations of $\lambda m_{\varphi}$ and $\lambda^2\Lambda$ used above for the scalarized black hole. In Table \ref{Tbl:hyp} we give the black hole models at which the equation (\ref{master_eq}) loses hyperbolicity for the pure massive case, and for the self-interaction case. At first we have confirmed the black hole mass and horizon radius, reported in \cite{BlazquezSalcedo2020}, at which the perturbation equation loses hyperbolicity for $\beta = 6$. In our case, however, we are using $\beta = 12$ and the critical point is at lower black hole radius compared to the one in \cite{BlazquezSalcedo2020} (about half of it), therefore increasing $\beta$ moves the critical point towards models with smaller radius (lower mass). Concerning the effect that the scalar field mass has on the critical point at which the equation loses hyperbolicity -- the radius of the black hole increases with the increase of the scalar field mass. Similarly, for models with fixed scalar field mass, the increase of the self-interaction constant moves the critical point towards higher black hole radii.

\begin{table}
	\caption{ The mass and the radius of the scalarized black holes at which the perturbation equation \eqref{master_eq} loses hyperbolicity. The results are for pure massive sGB theory with coupling \eqref{eq:coupling_function_scal} and for massive sGB with self-interaction.}
	\label{Tbl:hyp}\textbf{}
	\begin{center}
		\begin{tabular}{ | l | l | l | l |}
			\hline
			$\lambda m_{\varphi}$ & $\lambda^2\Lambda$ & $r_H/\lambda$ & $M/\lambda$  \\ \hline
	
			0 & 0 & $0.10812$  & $0.086401$  \\ \hline
			0.1 & 0 & $0.11375$  & $0.090276$ \\ \hline
			0.5 & 0 & $0.1463$  & $0.11214$  \\ \hline
			0.8 & 0 & $0.1867$  & $0.13818$  \\ \hline
			0.5 & 1 & $0.149$  & $0.11395$  \\ \hline
			0.5 & 5 & $0.1599$  & $0.12121$  \\ \hline
			0.5 & 10 & $0.17425$  & $0.13063$  \\ \hline		
		\end{tabular}
	\end{center}
\end{table}

\section{Conclusion}

Using gravitational waves observations to properly understand the nature of gravity in the strong field regime calls for a thorough study of quasinormal modes of compact objects in alternative theories. This motivates us to study the axial perturbations of black holes in a natural extension of GR such as the extended scalar-tensor theories of gravity, and more precisely in Gauss-Bonnet theory of gravity endowed with a self-interacting massive scalar field.

In the present paper we examined the axial QNMs of black holes for two coupling functions via time evolution of the time-dependent perturbation equation. We studied two types of coupling functions -- the first one being a linear function of the scalar field and the second one allowing for curvature induced spontaneous scalarization. In order to examine in detail the effect of nonzero scalar field potential we considered several sequences of models with fixed black hole mass and a range of values for the  scalar field mass and the self-interacting constant. 

First we studied black holes in a shift-symmetric Gauss-Bonnet theory with a linear coupling function that is often referred to as Einstein-dilaton Gauss-Bonnet gravity. For models with no self-interactions the Schwarzschild oscillation frequency (damping time) is highest (longest) with the rest of the models having slightly smaller (shorter) frequency (damping time). The deviations, however, are small. For models with fixed value for the mass of the scalar field and different values for the self-interaction constant the effect of the self-interaction constant is qualitatively the same as the effect of the mass of the scalar field. This is natural since both of them suppress the scalar field around the black hole.

For the scalarized black holes in a $Z_2$ symmetric scalar-Gauss-Bonnet gravity we studied models with two different fixed masses -- one higher mass $M/\lambda = 0.418$ close to the bifurcation point where new scalarized solutions appear, and a second smaller black hole mass  $M/\lambda = 0.3$  located far away from this bifurcation. For all cases the deviations in the oscillation frequencies are  small, with the Schwarzschild one being the lowest. Slightly higher deviations and more interesting behavior we observed for the damping times. For models with no self-interaction and $M/\lambda = 0.3$ the shortest damping time were found to be for the Schwarzschild black hole, followed by the case with zero scalar field potential, and it increases with the mass of the scalar field. For $M/\lambda = 0.418$ the shortest damping time is for the model with the heaviest scalar field (it is slightly shorter than the Schwarschild one), and for the rest of the models the damping times are higher than that and similar to one another. Our study of models with in-between masses show that the transition between the two described behaviors happens gradually with the increase of the black hole mass. For models with fixed mass of the scalar field and different values of the self-interaction constant we found the effect of the self-interaction constant to be qualitatively the same as the effect of the mass of the field.

Even though the effect of nonzero scalar field potential can have small effect on the QNM frequencies in certain cases, it has an important influence on the hyperbolicity of the system of perturbation equations. That is why we have examined in detail the effect of the scalar field mass and self-interaction constant on the critical point at which the perturbation equation change their character from  hyperbolic to mixed one. For both $\lambda m_{\varphi}$ and $\lambda^2 \Lambda$ when they increase, the critical point moves along the branch towards black hole models with larger radius which effectively shortens the range of stable black hole models. This observation can have a dramatic effect on the dynamical simulation involving scalarized black holes such as binary black hole mergers  and stellar core-collapse.  The results obtained in sGB gravity with vanishing scalar field potential demonstrated that the scalar field should be limited to low values if one wants to avoid loss of hyperbolicity of the fully nonlinear coupled equations \cite{Ripley:2020vpk,East:2021bqk,Kuan:2021lol}. The results in the present paper suggest that these limits should be even more severe for massive or self-interacting scalar field. This is important to be further explored because nonzero scalar field mass is often a desired feature of a theory admitting scalarization since it can ``hide'' effects such as the scalar dipole radiation and thus reconcile a theory with observations (e.g. the observations of orbital decay of double pulsars in close binary systems \cite{Damour1996,Popchev2015,Ramazanoglu2016,Yazadjiev2016}).

\section*{Acknowledgements}
KS and SY acknowledge financial support by the Bulgarian NSF Grant KP-06-H28/7. JK and JLBS gratefully acknowledge support from the DFG Research Training Group 1620 {\sl Models of Gravity} and DFG project BL 1553. JLBS would like to acknowledge support from projects PTDC/FIS-OUT/28407/2017 and PTDC/FIS-AST/3041/2020. DD acknowledge financial support via an Emmy Noether Research Group funded by the German Research
Foundation (DFG) under grant no. DO 1771/1-1.  SY would like to thank the University of T\"ubingen for the financial
support. Networking support by the COST actions CA15117 and CA16104 is gratefully acknowledged.


\bibliography{references}

\end{document}